\documentclass[aps,prl,floats, twocolumn,showpacs,superscriptaddress,reprint]{revtex4-1}
\usepackage{graphicx,epsfig}
\usepackage{graphics,dcolumn,bm,epic, eepic,fleqn,float}
\usepackage{amssymb,amsmath,amsfonts,multirow,rotate,color}
\usepackage{soul}
\definecolor{Myorange}{cmyk}{0,0.42,1,0}

\newcommand{\avg}[1]{\langle #1 \rangle}

\newcommand{\ud}{\,\mathrm{d}}

\begin{document}
\title{Remote synchronization reveals network symmetries and functional modules}

\author{Vincenzo Nicosia}
\affiliation{School of Mathematical Sciences, Queen Mary University of London, 
London E1 4NS, United Kingdom}  
\author{Miguel Valencia}
\affiliation{Neurophysiology Laboratory, CIMA,  University of Navarra, 31008 Pamplona, Spain}
\author{Mario Chavez}
\affiliation{CNRS UMR-7225, H\^{o}pital de la Salp\^{e}tri\`{e}re. 47 Boulevard de l'H\^{o}pital, 75013 Paris, France}
\author{Albert D\'iaz-Guilera} 
\affiliation{Dept. de F\'{\i}sica Fonamental, Facultat de F\'{\i}sica, Universitat de Barcelona, E-08028 Spain}
\author{Vito Latora}
\affiliation{School of Mathematical Sciences, Queen Mary University of London, 
London E1 4NS, United Kingdom}  
\affiliation{Dipartimento di Fisica ed Astronomia, Universit\`a di Catania and INFN, I-95123 Catania, Italy}


\begin{abstract}
  We study a Kuramoto model in which the oscillators are associated
  with the nodes of a complex network and the interactions include a
  phase frustration, thus preventing full synchronization. The system
  organizes into a regime of remote synchronization where pairs of
  nodes with the same network symmetry are fully synchronized, despite
  their distance on the graph. We provide analytical arguments to
  explain this result and we show how the frustration parameter
  affects the distribution of phases.  An application to brain
  networks suggests that anatomical symmetry plays a role in neural
  synchronization by determining correlated functional modules across
  distant locations.
\end{abstract}


\maketitle

Synchronization of coupled dynamical units is a ubiquitous phenomenon
in nature~\cite{piko}. Remarkable examples include phase locking in
laser arrays, rhythms of flashing fireflies, wave propagation in the
heart, and also normal and abnormal correlations in the activity of
different regions of the human
brain~\cite{strogatz,bullmore,chavez_brain,Tass1999}. In 1975
Y. Kuramoto proposed a simple microscopic model to study collective
behaviors in large populations of interacting
elements~\cite{kuramoto}.  In its original formulation the Kuramoto
model describes each unit of the system as an oscillator which
continuously readjusts its frequency in order to minimize the
difference between its phase and the phase of all the other
oscillators. This model has shown very successful in understanding the
spontaneous emergence of synchronization and, over the years, many
variations have been
considered~\cite{sakaguchi,strogatzrev_kuramoto,acebron}.
Recently, the Kuramoto model has been also extended to sets of
oscillators coupled through complex
networks~\cite{strogatz,watts,rev:bocc}, and it has been found that
the topology of the interaction network has a fundamental role in the
emergence and stability of synchronized
states~\cite{Wiley2006,PRsync}.  In particular, the presence of
communities ---groups of tightly connected nodes--- has a relevant
impact on the path to
synchronization~\cite{arenas_modules,bocca_modules,gardenes,adaptive,luce},
and units that are close to each other on the network, or belong to
the same module or community~\cite{fortunato}, have a higher chance to
exhibit similar dynamics. This implies that nodes in the same
structural module share similar functions, which is a belief often
supported by empirical findings~\cite{Tegner2003,bullmore}.  However,
various examples are found in nature where functional similarity is
instead associated with morphological symmetry. In these cases, units
with similar roles, which could potentially swap their position
without altering the overall functioning of the system, appear in
remote locations of the network. Some examples include cortical areas
in brains~\cite{varela2001}, symmetric organs in plants and
vertebrates~\cite{Smith2006,Ruvinsky2000}, and even atoms in complex
molecules~\cite{Rosenthal1936}. Therefore, identifying the sets of
symmetric units of a complex system might be helpful to understand its
organization.
Finding the global symmetries in a graph, i.e., constructing its
automorphism group, is a classical problem in graph theory. However,
it is still unknown if this problem is polynomial or
NP-complete~\cite{footnote1,Papadimitriou1994}, even if there exist
polynomial-time algorithms for graphs with bounded maximum
degree~\cite{luks82}. Recent works have focused instead on defining
and detecting local symmetries in complex
networks~\cite{holme1,holme2}. Nevertheless, the interplay between the
structural symmetries of a network and the dynamics of processes
occurring over the network has been studied only
marginally~\cite{Russo2011,remote,Aufderheide2012}, or for specific
small network motifs~\cite{Fischer2006,Vicente2008,Viriyopase2012}.

In this Letter we show that network symmetries play a central role in
the synchronization of a system. We consider networks of identical
Kuramoto oscillators, in which a phase frustration parameter forces
connected nodes to maintain a finite phase difference, thus hindering
the attainment of full synchronization. We prove that the
configuration of phases at the synchronized state reflects the
symmetries of the underlying coupling network. In particular, two
nodes with the same symmetry have identical phases, i.e., are fully
synchronized, despite the distance between the two nodes on the
graph. Such a remote synchronization behavior is here induced by the
network symmetries and not by an initial \textit{ad hoc} choice of
different natural frequencies~\cite{remote}.

Let us consider $N$ identical oscillators associated to the nodes of a
connected graph $G({\cal N}, {\cal L})$, with $N=| {\cal N} |$ nodes
and $K= | {\cal L} |$ links. Each node $i$ is characterized, at time
$t$, by a phase $\theta_i(t)$ whose time evolution is governed by the
equation
\begin{equation}
 \dot{\theta}_i =\omega + \lambda \sum_{j = 1}^N a_{ij} 
\sin ( \theta_j -\theta_i  -\alpha )\;.
\label{kura}
\end{equation}
Here $\omega$ is the natural frequency, identical for all the
oscillators, and $A \equiv \{a_{ij}\}$ is the adjacency matrix of the
coupling graph. The model has two control parameters: $\lambda>0$
accounting for the strength of the interaction, and $\alpha$, the
phase frustration parameter ranging in $[0, \pi/2]$.
When $\alpha=0$, the model reduces to a network of identical Kuramoto
oscillators.  In this case, the fully synchronized state is globally
stable for a set of initial conditions having finite
measure~\cite{kuramoto,acebron} and the transient dynamics closely
reflects the structure of the graph, so that nodes belonging to the
same structural module evolve similarly in time~\cite{arenas_modules}.
However, the synchronized state can coexist with other nontrivial
attractors, e.g., uniformly twisted waves, especially if the coupling
topology is regular and sparse (see Ref.~\cite{Wiley2006} for a
discussion about the size of the sync basin).
Instead, if the oscillators are not identical the frequency
distribution tends to separate their phases and, as a result, there is
a transition from an incoherent state (with order parameter
$r=\frac{1}{N} \left| \sum_{j=1}^{N}{\mbox e}^{i\theta_j}\right| $
equal to 0) to a synchronized one ($r \neq 0$) at a critical value
$\lambda_c$ of the coupling strength.

\begin{figure}[b]
\centering
\includegraphics[width=3in,angle=-0]{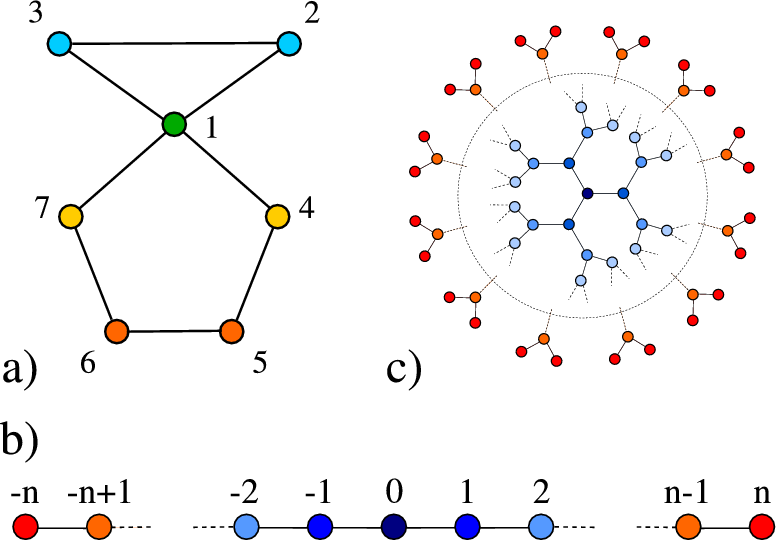}
\caption{(color online) The presence of frustration reveals clusters
  of symmetric nodes. The color code represents the phases of nodes at
  a given time in the stationary state. \textbf{(a)} In the first
  graph ($G_a$), node 2 is synchronized to node 3, node 4 to node 7,
  and node 5 to node 6. \textbf{(b)} In a finite chain ($G_b$), pairs
  of nodes symmetrically placed with respect to the central node are
  perfectly synchronized. \textbf{(c)} In a finite Bethe lattice
  ($G_c$) all the nodes placed at the same distance from the center
  have equal phases.}
\label{fig1}
\end{figure}

\begin{figure}[t]
\centering \includegraphics[width=3in]{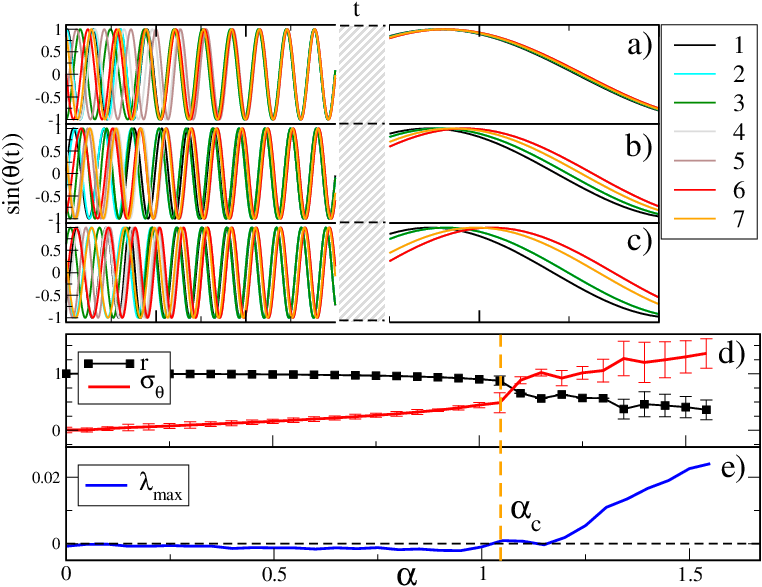}
\caption{(color online) The figure refers to the coupling topology
  $G_a$ in Fig.~\ref{fig1}a. Panels (a)--(c): after an initial
  transient the system reaches a phase-locked synchronized state in
  which symmetric nodes have the same phase.  The panels correspond to
  three different values of the frustration parameter, respectively,
  (a) $\alpha=0.1$, (b) $\alpha=0.5$, (c) $\alpha=0.8$. Panel (d): for
  $\alpha>\alpha_c$ the synchronized state becomes unstable, the order
  parameter decreases while the dispersion of the phases
  $\sigma_{\theta}$ increases. Panel (e): the maximum Lyapunov
  exponent $\lambda_{max}$ of oscillators coupled through $G_a$
  becomes positive for $\alpha>\alpha_c$, and the systems enters a
  chaotic regime. The dashed yellow line indicates the approximate
  position of $\alpha_c$.}
\label{fig2}
\end{figure}

The introduction of a phase frustration $\alpha \neq 0$ forces
directly connected oscillators to maintain a constant phase difference
\cite{nota}.  In particular, we found that for a wide range of
$\alpha>0$ the dynamics in Eqs.~(\ref{kura}) reaches a stationary
state in which the oscillators at two symmetric nodes have exactly the
same phase, and this phase differs from the phases of nodes with
different symmetries.  Let us first illustrate this behavior and the
effect of $\alpha$ on the three graphs $G_a$, $G_b$ and $G_c$ shown in
Fig.~\ref{fig1}.
In the three topmost panels of Fig.~\ref{fig2} we report the results
of the numerical integration of Eqs.~(\ref{kura}) on the graph $G_a$
for three different values of $\alpha$.  We find that, after a
transient, the system settles into a stationary state in which, at any
time $t$, the phases are grouped into four different trajectories:
$\theta_1(t)$, $\theta_2(t)= \theta_3(t)$, $\theta_4(t)= \theta_7(t)$
and $\theta_5(t)=\theta_6(t)$. In general, by increasing $\alpha$ up
to a certain value $\alpha_c$ we better separate the four
trajectories.

The four clusters of nodes obtained for $\alpha<\alpha_c$ are
identified by a color code in Fig.~\ref{fig1}. We notice that each
cluster groups together all the nodes with the same symmetry. In this
way two distant nodes of the graph, e.g. node $4$ and node $7$, are
fully synchronized even if the other nodes in the paths connecting
them have different phases. In this respect, what we observe is a
remote synchronization~\cite{remote}. We have found similar results
for the linear chain and for the Bethe lattice (see nodes with the
same colors in $G_b$ and $G_c$ in Fig.~\ref{fig1}).

Notice that if the system reaches a synchronized state and $\alpha$ is
small enough, Eq.~(\ref{kura}) can be linearized by replacing the
sinus with its argument. We obtain
\begin{equation}
 \dot{\theta}_i = \omega - \lambda 
\left[  \sum_{j = 1}^N L_{ij} \theta_j + \alpha k_i  \right] 
~~ 
\label{kura_lin}
\end{equation}
where $k_i=\sum_j a_{ij}$ is the degree of node $i$, $L_{ij}$ are the
entries of the Laplacian matrix of the graph $L \equiv D - A$, and $D$
is a diagonal matrix such that $D_{ii}=k_i$.  Without loss of
generality, we can set $\lambda=1$, $\omega=0$. If the system is
synchronized then $\dot{\theta_i} = \Omega, \forall i$, so that the
phases must satisfy the equations $\sum_{j = 1}^N L_{ij} \theta_j =
\alpha \left[ \avg{k} - k_i \right]$ at any time, or equivalently
\begin{equation}
  L \bm{\theta} =  \alpha \left[\avg{k}\bm{1} - \bm{k} \right]
  \label{kura_lin_equi}
\end{equation}
where $\avg{k}=N^{-1} \sum_{i}k_i$ is the average degree of the
network. This corresponds to a synchronization frequency
$\dot{\theta}_i\!=\!\Omega\!=\!-\alpha\avg{k}\>\forall i$. In a
connected graph the Laplacian matrix has one null eigenvalue and the
system of Eqs.~(\ref{kura_lin_equi}) is singular. Consequently, at
each time $t$ we can solve the system by computing the phase
difference between each node and a given node chosen as reference.
For instance, if in $G_a$ we define $\phi_j(t) = \theta_j(t) -
\theta_1(t), j=2,\ldots,7$, by solving Eqs.~(\ref{kura_lin_equi}) we
obtain $\phi_2 = \phi_3 = \alpha\left[\avg{k} - 2\right]$, $\phi_4 =
\phi_7 = 2\alpha\left[\avg{k} - 2\right]$, and $\phi_5 = \phi_6 =
3\alpha\left[\avg{k} - 2\right]$.
This is in agreement with the results of the simulations: the phases
are clustered into four groups, with nodes with the same symmetry
having the same phase, and nodes with different symmetries being
separated by a phase lag that depends on $\alpha$ as in the relations
found above. An analogous analytical expression can be derived for a
finite chain (graph $G_b$ in Fig.\ref{fig1}), for which we obtain
$\theta_n-\theta_{n-i}=\theta_{-n}-\theta_{-n+i}=
\left[\frac{i(i+1)}{2}\avg{k}-i^2 \right]\alpha $ and
$\theta_{n}\!-\!\theta_0\!=\!\theta_{-n}\!-\!\theta_{0}\!=\!\left[\frac{n(n+1)}{2}\avg{k}
  - n^2 \right]\alpha$. Consequently, two nodes symmetrically placed
with respect to node $0$ will have identical phases. 

We now provide a general argument to explain why the synchronization
of Eqs.~(\ref{kura}) is related to graph symmetries.  A graph $G({\cal
  N}, {\cal L})$ has a symmetry if and only if it is possible to find
a bijection $\pi: {\cal N} \rightarrow {\cal N}$ which preserves the
adjacency relation of $G$, i.e., which is an automorphism for
$G$. Formally, this means that there exists a permutation matrix
$P=P(\pi)$ such that $PAP^{-1} = A$. If $P$ corresponds to an
automorphism of $G$ then $P$ commutes with $A$, i.e., $PA=AP$, and
$PAP^{-1}$ performs a relabeling of the nodes of the original graph
which preserves the adjacency matrix~\cite{footnote2}.  In general a
graph can admit more than one automorphism. For instance, graph $G_a$
in Fig.\ref{fig1} has at least three nontrivial bijections which
preserve the adjacency matrix, namely
$$
\begin{array}{llcr}
\pi_{1}& : (1,2,3,4,5,6,7) & \rightarrow & (1,3,2,4,5,6,7)\\
\pi_{2}& : (1,2,3,4,5,6,7) & \rightarrow & (1,2,3,7,6,5,4)\\
\pi_{3}& : (1,2,3,4,5,6,7) & \rightarrow & (1,3,2,7,6,5,4)\\
\end{array}
$$
\begin{figure}[b]
  \begin{center}
    \includegraphics[width=3.1in]{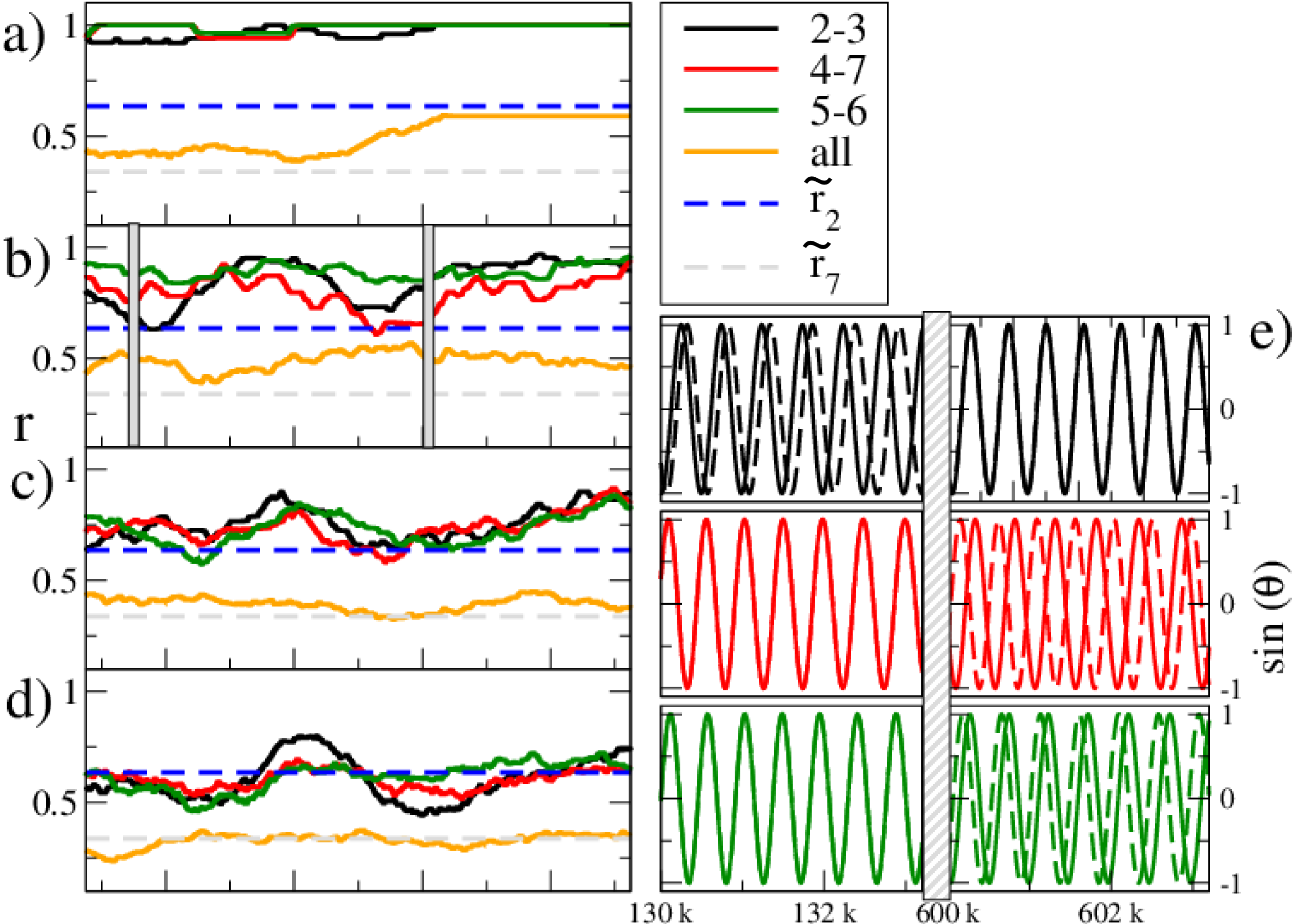}
  \end{center}
  \caption{(color online) Chaotic regime in $G_a$ when
    $\alpha>\alpha_c$. Panel (a)--(d): running averages of $r$
    (orange) and pairwise order parameters $r_2$ (black, red and green
    lines) for typical trajectories of oscillators coupled through
    graph $G_a$, when $\alpha$ is respectively equal to (a) $1.3$, (b)
    $1.4$, (c) $1.5$ and (d) $1.55$. The dashed lines indicate the
    expected synchronization level for a system of two (blue line,
    $\tilde{r_2} = 2/\pi$) and seven incoherent oscillators (gray
    line, $\tilde{r_7}\simeq 0.338\ldots$).  Panel (e) the plot of the
    phases of pairs of symmetric nodes for $\alpha=1.4$ in two
    different temporal intervals (the shaded gray regions in panel b)
    reveal the existence of metastable, partially synchronized
    states.}
  \label{fig3}
\end{figure}
Node $2$ and node $3$ are symmetric because we can relabel the nodes
of $G_a$ (e.g. by means of either $\pi_1$ or $\pi_3$) so that node $2$
is mapped into node $3$ and vice versa, and the adjacency matrix of
$G_a$ is left unchanged. Similarly, for the pairs $\{4,7\}$ and
$\{5,6\}$, there are two different relabelings which preserve
adjacency relations, i.e., $\pi_2$ and $\pi_3$.  In terms of
symmetries, the graph $G$ has the following four different classes of
nodes: $C_1=\{ 1 \}$, $C_2=\{ 2,3 \}$, $C_3=\{ 4,7 \}$, $C_4=\{ 5,6
\}$.  Now, if a permutation of the nodes is an automorphism of $G$,
then $PLP^{-1} = PDP^{-1} - PAP^{-1} = D - A = L$, i.e., the
associated permutation matrix $P$ also commutes with the Laplacian
matrix of the graph. By left-multiplying both sides of
Eq.~(\ref{kura_lin_equi}) by $P$, we get $PL\bm{\theta} = \alpha
P\left[\avg{k}\bm{1} -\bm{k} \right]$.
Since $PL=LP$ ($P$ commutes with $L$) and $P\bm{k} = \bm{k}$
(symmetric nodes have the same degree) then we have
\begin{equation}
  LP\bm{\theta} = \alpha\left[\avg{k}\bm{1} -\bm{k}\right]
\label{eq:final}
\end{equation}
Combining Eq.~(\ref{kura_lin_equi}) and Eq.~(\ref{eq:final}) we
finally obtain the linear system
\begin{equation}
  LP\bm{\theta}=L\bm{\theta}
\end{equation}
which is singular, i.e., has one free variable. Again, it can be solved
by leaving free one of the $N$ variables $\theta_i$, setting $\phi_j =
\theta_j - \theta_i$ and considering the new system
$\tilde{L}\tilde{P}\bm{\phi}=\tilde{L}\bm{\phi}$. The matrix
$\tilde{P}$ is obtained from $P$ by removing the row and the column
corresponding to node $i$. If $P$ does not permute node $i$ with
another node, then $\tilde{P}$ is still a permutation matrix.
Similarly, $\tilde{L}$ is the reduced Laplacian, i.e., the matrix
obtained from the Laplacian by deleting the $i-$th row and the $i-$th
column.  By left-multiplying by $\tilde{L}^{-1}$, which is not
singular, we obtain
\begin{equation}
  \tilde{P}\bm{\phi} = \bm{\phi}
  \label{eq:phase_perm}
\end{equation}
Since $\tilde{P}\bm{\phi}$ is a permutation of the phases of symmetric
nodes, Eq.(\ref{eq:phase_perm}) implies that the phases of symmetric
nodes will be equal at any time, whereas by solving
Eq.~(\ref{eq:final}) we can get the values of the corresponding
phases.
This argument is valid for small values of $\alpha$, since the
linearization of Eq.~(\ref{kura}) is possible only if
$\sin(x-\alpha)\simeq (x-\alpha)$, but as shown in
Fig.~\ref{fig2}(a)-(c) we observe the formation of the same perfectly
synchronized clusters of symmetric nodes for a wide range of $\alpha$.
However, when $\alpha$ becomes larger than a certain value $\alpha_c$,
the assumption $\dot{\theta_i} = \Omega, \forall i$ does not hold any
more and the global synchronized state loses stability. By looking at
Fig.~\ref{fig2}(d)-(e) we notice that for $\alpha>\alpha_{c}$, with
$\alpha_c\simeq 1.05$ for the graph $G_a$, the value of $r$ steadily
decreases while the dispersion of phases increases, until it reaches
the expected value $\bar{\sigma}_{\theta}\simeq1.39$ for a system of
seven incoherent oscillators (see Fig.~\ref{fig2}(d) and
Appendix). Moreover, for $\alpha > \alpha_c$ the maximal Lyapunov
exponent of the system $\lambda_{max}$ becomes positive and the system
enters a chaotic regime (see Fig.~\ref{fig2}(e)). Interestingly, the
results reported in Fig.~\ref{fig3}(a)-(d) confirm that in this regime
the coherence of symmetric nodes, measured by the pairwise order
parameter $r_2$, is higher than expected for incoherent oscillators
(refer to Appendix for additional details). Figure~\ref{fig3}(e) shows
that for $\alpha>\alpha_c$ the system exhibits metastable, partially
synchronized states, in which pairs of symmetric nodes alternates
intervals of perfect synchronization with intervals of complete
incoherence. We point out that in this regime chimera states could
potentially occur~\cite{chimera1,Laing2009,Shanahan2010,Martens2010}
and could even coexist with remote synchronization for
$\alpha<\alpha_c$. Qualitatively similar results are obtained for
different coupling topologies, but the actual value of $\alpha_c$
seems to depend on the structure of the coupling network in a
nontrivial way.

{\em Application to the brain.--} As an example, we investigate here
the role of symmetry in the human brain by considering anatomical and
functional brain connectivity graphs defined on the same set of $N=90$
cortical areas (see details in Appendix). We have first constructed a
graph of anatomical brain connectivity as obtained from DW-MRI
data~\cite{iturria2011}, where links represent axonal fibers, and we
used this graph as a backbone network to integrate Eqs.~(\ref{kura}).
\begin{figure}[t]
\centering
\includegraphics[width=3in]{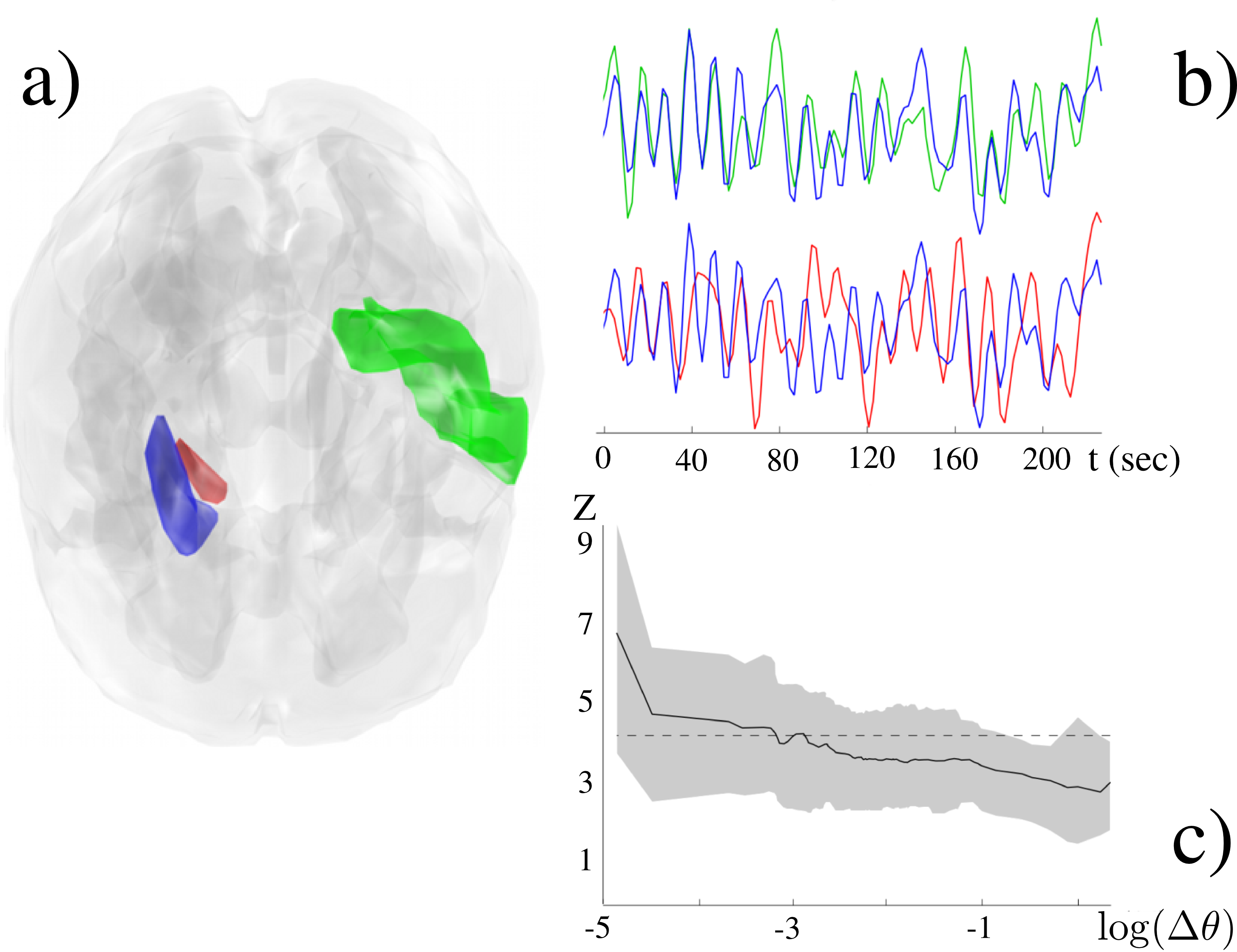}
\caption{(color online) (a) Brain areas with similar and dissimilar
  phases of the frustrated Kuramoto model are colored and superimposed
  onto an anatomical image. (b) Examples of functional data from one
  subject recorded at the brain areas indicated in panel (a). Colors
  are the same as those used in the anatomical image. (c) Functional
  correlation $Z$ between pairs of nodes as a function of their phase
  differences $\Delta \theta$ according to the simulated Kuramoto
  dynamics. The black solid curve corresponds to the average value
  over all the subjects, while the gray area covers the
  $5^{\text{th}}$ and the $95^{\text{th}}$ percentiles of the
  distribution. The dashed horizontal line indicates the threshold for
  statistical significant correlations ($p<0.05$, corrected for
  multiple comparisons).}
\label{fig4}
\end{figure}
We identified candidate pairs of anatomically symmetric areas by means
of agglomerative clustering, i.e., grouping together nodes having
close phases at the stationary state (full dendrogram and details are
provided in Appendix).  Then, we considered the graph of functional
brain connectivity, in which links represent statistically significant
correlations between the BOLD fMRI time-series of cortical areas (see
details in Appendix). Figure~\ref{fig4} illustrates the results for
$\alpha=0.5$ (we obtained qualitatively similar results in a wide
range of $\alpha$). Consider nodes 57 and 74, corresponding
respectively to the green and blue areas in panel (a). Not only the
two areas are spatially separated, but there is no edge connecting the
two corresponding nodes in the anatomical connectivity
network. However, the two nodes are detected as a candidate symmetric
pair since at the stationary state of the Kuramoto dynamics in
Eq.~(\ref{kura}) the oscillators associated to these two nodes have
very close phases (see dendrogram in Appendix). As shown in
Fig.~\ref{fig4}(b), also the BOLD fMRI signals corresponding to nodes
57 and 74 also are strongly synchronized. We obtain remarkably
different results when we consider node 74 and node 76. These nodes
correspond to two spatially adjacent areas of the brain (the red and
blue regions in Fig.~\ref{fig4}(a) and are directly connected in the
anatomical connectivity network. However, at the stationary state of
Eq.~(\ref{kura}) the phase difference of the oscillators associated to
node 74 and 76 is quite large. Interestingly, in this case the fMRI
time-series associated to these nodes are much less similar to each
other [see the two bottom trajectories reported in
  Fig.~\ref{fig4}(b)].

To quantify this effect, we plot in Fig.~\ref{fig4}c the average
functional correlation $Z$ between the fMRI activity of pairs of brain
areas as a function of the phase differences $\Delta \theta$ between
the phases of the corresponding oscillators, obtained from the
dynamics of Eq.~(\ref{kura}) on the anatomical connectivity
network. The fact that $Z$ decreases with $\Delta \theta$ suggests
that structural symmetry plays an important role in determining human
brain functions. In fact, the functional activities of anatomically
symmetric areas can be strongly correlated, even if the areas are
distant in space. These results suggest that the study of anatomical
symmetries in neural systems might provide meaningful insights about
the functional organization of distant neural assemblies during
diverse cognitive or pathological states~\cite{varela2001}. Applied to
other connectivity networks as a method to spot potential network
symmetries, our study could provide new insights on the interplay
between structure and dynamics in complex systems.

\medskip
\begin{acknowledgments}
The authors thank Yasser Iturria-Medina for sharing the DTI
connectivity data used in the study, and Simone Severini for useful
comments.  M. V. acknowledges financial support from the Spanish
Ministry of Science and Innovation, Juan de la Cierva Programme
Ref. JCI-2010-07876. A. D-G. acknowledges support from the Spanish
DGICyT Grant FIS2009-13730, from the Generalitat de Catalunya
2009SGR00838.  This work was supported by the EU-LASAGNE Project,
Contract No.318132 (STREP).
\end{acknowledgments}

\renewcommand\theequation{{S-\arabic{equation}}}
\renewcommand\thetable{{S-\Roman{table}}}
\renewcommand\thefigure{{S-\arabic{figure}}}
\setcounter{equation}{0}
\setcounter{figure}{0}
\setcounter{section}{0}

\section*{Appendix}

\noindent
\textbf{Expected order parameter for two incoherent oscillators. --}
In Fig.3 of the main text we showed that when $\alpha$ approaches
$\pi/2$ the order parameter $r_2$ for a pair of symmetric oscillators
remains higher than the expected value $\bar{r}_2$ for a pair of
incoherent oscillators. We report here the derivation of $\bar{r}_2$
for a pair of oscillators that are not synchronized. We assume that
the phases $\theta_1$ and $\theta_2$ of the two oscillators are drawn
uniformly in $[0,2\pi]$. Thanks to the rotational symmetry, we can set
one of the phases equal to $0$, e.g. $\theta_1=0$, while drawing the
other uniformly in $[0,2\pi]$. We notice that the value of $r_2$ for
the generic pair of phases $(\theta_1=0,\theta_2)$ reads:
\begin{eqnarray*}
  r_2 & = & \frac{1}{2}\sqrt{\left(\cos{\theta_1} +
    \cos{\theta_2}\right)^2 + \left(\sin{\theta_1} +
    \sin{\theta_2}\right)^2} =\\ & = & \frac{1}{2}\sqrt{\left(1 +
    \cos{\theta_2}\right)^2 + \sin^2{\theta_2}} =\\ & = &
  \frac{1}{2}\sqrt{2 + 2 \cos{\theta_2}}
\end{eqnarray*}
Consequently, the expected value of $r_2$ is obtained as the average
over all the possible choices of $\theta_2$, namely:
\begin{eqnarray*}
  \tilde{r_2} & = & \frac{1}{4\pi}\int_{-\pi}^{\pi} \sqrt{2 + 2
    \cos{\theta}} \ud \theta = \\ & = &
  \frac{1}{\pi}\int_{-\pi/2}^{\pi/2}\cos{u}\ud u = \\ & = &
  \frac{2}{\pi} \simeq 0.6366197\ldots
\end{eqnarray*}

\bigskip

\noindent
\textbf{Phase dispersion and numerical estimate for incoherent
  oscillators. --} In Fig.2d of the main text we reported, as a
function of $\alpha$, the dispersion of the phases $\sigma_{\theta}$
for a system of seven oscillators coupled through graph $G_a$ in
Fig.1a. We noticed that the dispersion approaches the value $1.39$
when $\alpha\rightarrow\pi/2$. Given a set of $N$ unitary vectors
having phases $\{\theta_1, \theta_2,\ldots,\theta_N\}$, consider the
average vector:
\begin{equation}
  r e^{\text{i}\psi} = \frac{1}{N}\sum_{j=1}^{N}e^{\text{i}\theta_{j}}
\end{equation}
having polar coordinates $(r,\psi)$. The dispersion of the $N$ phases
of the set around the phase $\psi$ of the average vector is defined
as:
\begin{equation}
  \sigma_{\theta} = \sqrt{\frac{1}{N} \sum_{j=1}^{N}(\theta_j - \psi)^2}
\end{equation}
where the difference $(\theta_j - \psi$) is computed $\text{mod}(2\pi)$
and takes in $[0,\pi]$. An estimate of the expected phase dispersion
for a set of $N$ incoherent oscillators can be obtained as the average
over $M$ independent realizations of $\sigma_{\theta}$ computed on a
set of $N$ phases uniformly drawn in $[0,2\pi]$.  For the system of
$N=7$ oscillators coupled through graph $G_a$ we averaged
$\sigma_{\theta}$ over $M=10^7$ samples, obtaining the estimate
$\bar{\sigma}_{\theta}=1.394 \pm 0.248$.

\bigskip
\noindent
\textbf{Computation of the maximum Lyapunov exponent
  $\lambda_{max}$. --} The computation of the maximum Lyapunov
exponent for the system coupled through graph $G_a$ of Fig.~(1) in the
main text was performed using equally separated values of $\alpha$
between $0$ and $1.57\simeq\pi/2$ ($\Delta\alpha=0.01$). Then, for
each value of $\alpha$, we considered 500 different initial
configurations of the phases of the seven oscillators. For each
initial condition, we let evolve the trajectory $\bm{\theta}(t)$
according to Eq.~(1) in the main text, until it reached the attractor
(or the stationary state, for $\alpha\ll \pi/2$). To properly take
into account the rotational symmetry of the system, we studied the
evolution of $\bm{\theta}(t)$ in Cartesian coordinates, i.e., by
looking at the set of $2N$ variables $\bm{x} =\{\cos{\theta_1},
\cos{\theta_2}, \ldots, \cos{\theta_N}, \sin{\theta_1},
\sin{\theta_2}, \ldots, \sin{\theta_N}\}$. We considered a
perturbation of $\bm{x}$ of magnitude $\varepsilon=10^{-4}$, i.e., a
trajectory $\tilde{\bm{x}}$ such that $d(\bm{x}(t), \tilde{\bm{x}}(t))
= \varepsilon$.  Here $d(\cdot, \cdot)$ denotes the Euclidean distance
in $\mathbb{R}^{2N}$. Then, we integrated both trajectories for one
integration step $h$ (using a standard fourth-order Runge-Kutta
integration scheme), we measured the distance $d_1 = d(\bm{x}(t+h),
\tilde{\bm{x}}(t+h))$ and we computed $\lambda =
\log{(d_1/\varepsilon)}$. The quantity $\lambda$ is a one-step
approximation of the largest Lyapunov exponent of the
system~\cite{Strogatz1994a}. Then, we realigned the perturbed
trajectory $\tilde{\bm{x}}$ so that the distance between $\bm{x}(t+h)$
and the realigned perturbed trajectory $\tilde{\tilde{\bm{x}}}(t+h)$
was equal to $\varepsilon$ in the same direction of $d_1$, and we
iterated the procedure. The value of $\lambda_{max}$ for a set of
initial conditions was obtained by averaging the values of $\lambda$
computed at each iteration over $10^4$ subsequent integration steps.

\begin{figure*}[!ht]
  \centering 
  \includegraphics[width=6in]{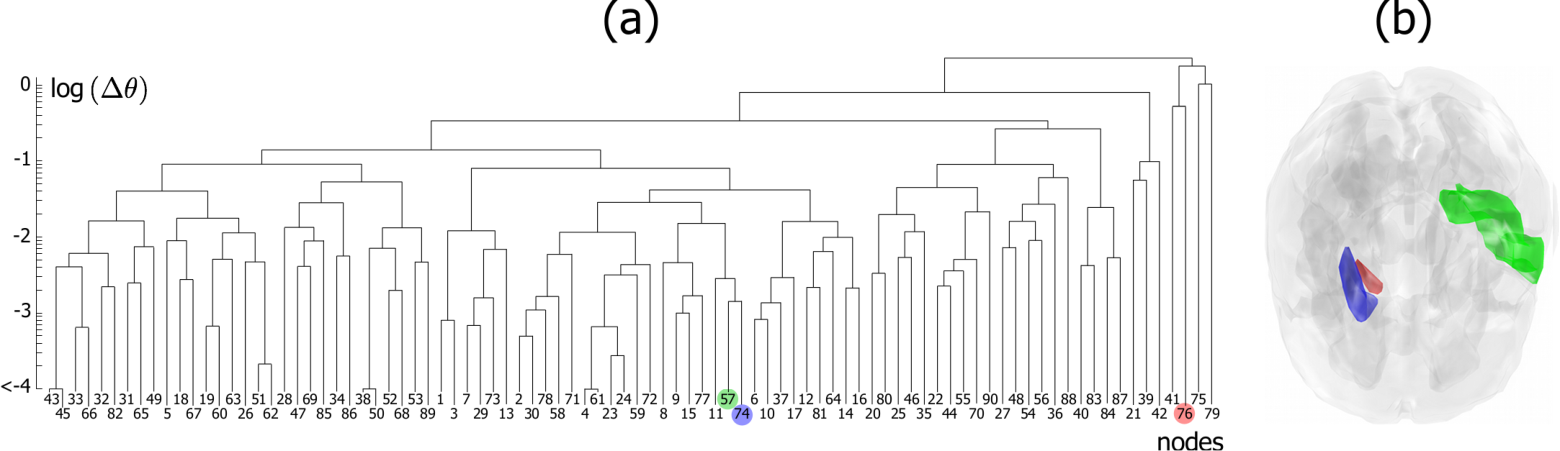}
    \caption{Hierarchical clustering of phase values obtained from
      anatomical connectivity in Eq.~(1) of the main text. Nodes
      correspond to the different brain areas. Brain sites with
      similar and dissimilar phases are colored and superimposed onto
      an anatomical image (see details in the main text).}
    \label{figure1SuppMat}
\end{figure*}

\bigskip
\noindent
\textbf{Brain data acquisition and pre-processing. --} The anatomical
connectivity network is based on the connectivity matrix obtained by
Diffusion Magnetic Resonance Imaging (DW-MRI) data from 20 healthy
participants, as described in~\cite{DTIdata2008}. The elements of this
matrix represent the probabilities of connection between the 90
anatomical regions of interest ($N=90$ nodes in the network) of the
Tzourio-Mazoyer brain atlas~\cite{TzourioMazoyer2002}. These
probabilities are proportional to the density of fibers between
different areas, so each element of the matrix represents an
approximation of the connection strength between the corresponding
pair of brain regions.

The functional brain connectivity was extracted from BOLD fMRI resting
state recordings obtained as described in~\cite{valencia2009}. All
fMRI data sets (segments of 5 minutes recorded from 15 healthy
subjects) were co-registered to the anatomical data set and normalized
to the standard MNI (Montreal Neurological Institute) template image,
to allow comparisons between subjects. As for DW-MRI data, normalized
and corrected functional scans were sub-sampled to the anatomical
labeled template of the human
brain~\cite{TzourioMazoyer2002}. Regional time series were estimated
for each individual by averaging the fMRI time series over all voxels
in each region (data were not spatially smoothed before regional
parcellation). To eliminate low frequency noise (e.g.~slow scanner
drifts) and higher frequency artifacts from cardiac and respiratory
oscillations, time-series were digitally filtered with a finite
impulse response (FIR) filter with zero-phase distortion (bandwidth
$0.01-0.1$~Hz) as in~\cite{valencia2009}.

\bigskip
\noindent
\textbf{Functional synchrony. --} A functional link between two time
series $x_i(t)$ and $x_j(t)$ (normalized to zero mean and unit
variance) was defined by means of the linear cross-correlation
coefficient computed as $r_{ij} = \langle x_i(t)x_j(t)\rangle $, where
$\langle\cdot\rangle$ denotes the temporal average. For the sake of
simplicity, we only considered here correlations at lag zero. To
determine the probability that correlation values are significantly
higher than what is expected from independent time series, $r_{ij}(0)$
values (denoted $r_{ij}$) were firstly transformed by the Fisher's Z
transform
\begin{equation}
Z_{ij} = 0.5\ln \left(\frac{1+r_{ij}}{1-r_{ij}} \right)
\end{equation}
Under the hypothesis of independence, $Z_{ij}$ has a normal
distribution with expected value 0 and variance $1/(df-3)$, where $df$
is the effective number of degrees of freedom~\cite{Bartlett1946,
  Bayley1946, Jenkins1968}. If the time series consist of independent
measurements, $df$ simply equals the sample size, $N$.  Nevertheless,
autocorrelated time series do not meet the assumption of independence
required by the standard significance test, yielding a greater Type I
error~\cite{Bartlett1946, Bayley1946, Jenkins1968}. In presence of
auto-correlated time series $df$ must be corrected by the following
approximation:
\begin{equation}
\frac{1}{df}\approx \frac{1}{N} + \frac{2}{N}\sum_\tau r_{ii}(\tau) r_{jj}(\tau), 
\end{equation}
where $r_{xx}(\tau)$ is the autocorrelation of signal $x$ at lag $\tau$. 

To estimate a threshold for statistically significant correlations,
a correction for multiple testing was used. The False Discovery Rate
(FDR) method was applied to each matrix of $Z_{ij}$
values~\cite{FDR}. With this approach, the threshold of significance
$Z_{\text{th}}$ was set such that the expected fraction of false
positives is restricted to $q \leq 0.05$.

\bigskip
\noindent
\textbf{Clustering of phase values. --} To identify brain areas that
could be related by a topological symmetry, we used the anatomical
connectivity obtained from the DW-MRI data as the connectivity matrix
($N=90$ nodes) in Eq.~(1) of the main text. A standard hierarchical
agglomerative clustering algorithm was then used to identify nodes
with similar phases~\cite{GanBook}. The resulting dendrogram is
depicted in Fig.~\ref{figure1SuppMat}.


\begin{thebibliography}{99}
\bibitem{nota} In general, Eqs.~(\ref{kura}) do not admit a fully
  synchronized stationary state, because the system
  $\Omega=\dot{\theta_i}=\omega-\lambda k_i\sin\alpha~\forall i$ has no
  solutions except for regular graphs, where $k_i=k~\forall i$.
\bibitem{footnote2} Each row and each column of a permutation matrix
  has exactly one entry equal to one and all others equal to $0$. If
  $P$ is a permutation matrix, the matrix product $PA$ swaps pairs of
  rows of $A$, while $AP^{-1}$ swaps pairs of columns of $A$.
\end{thebibliography}
\end{document}